\documentclass{article}
\input{tcilatex}

\begin{document}

\title{Conformation-networks of two-dimensional lattice homopolymers }
\author{Yu-Pin Luo$^{a}$, {Hung-Yeh Lin}$^{a}$, {Ming-Chang Huang}$^{a}$,
and Tsong-Ming Liaw$^{b}$ \\
%EndAName
$^{a}$Department of Physics, Chung-Yuan Christian University,Chungli
320,Taiwan\\
$^{b}$Computing Center, Academia Sinica, Nankang 115, Taiwan}
\maketitle

\begin{abstract}
The effect of different Monte Carlo move sets on the the folding kinetics of
lattice polymer chains is studied from the geometry of the
conformation-network. The networks have the characteristics of small- world.
The Monte Carlo move, rigid rotation, has drastic effect on the geometric
properties of the network. The move not only change the connections but also
reduce greatly the shortest path length between conformations. The networks
are as robust as random network.
\end{abstract}

Protein folding is a complex process for which, a sequence of amino acids
folds into a unique and stable structure in a relatively short time\cite%
{creigh}. The lattice models have been used widely as coarse-grained models
for the theoretical study of folding process\cite%
{skolnick,shakhnovich,miller,chan1,chan2,sali}. In the lattice models,
protein is viewed as a chain of $m$ monomers, and the conformations are
given by all possible self-avoiding walks of the chain on two-dimensional ($%
2D$) or three-dimensional $\left( 3D\right) $ regular lattices. The energy
of a conformation generally depends on the number of intrachain contacts,
and how to assign the contact energy is model dependent. The kinetics of the
folding process then is studied by Monte Carlo simulations for which, a move
set is designed for the change of conformations. In principle, different
move sets, satisfying the requirement of ergodicity, should reach the same
equilibrium canonical distribution after sufficiently long time simulations.
However, different move sets may yield different perspectives of folding
kinetics. The question of how different move sets affect folding kinetics
was discussed before by Chan and Dill\cite{chan1,chan2}. Based on two
different sets, they used transfer matrix with Metropolis criterion to study
the folding kinetics of two-dimensional homo- and hetero-polymers. The
results indicate that the kinetic sequence of folding events and the shape
of the energy landscape depend strongly on the move set. Hoang and Cieplak
also obtained the same conclusions after comparing the dynamics of three
different move sets\cite{hoang}. Therefore, it is important to understand
the nature of a move set adopted in the lattice dynamics. \newpage The
purpose of this Letter is to explore the characteristics of different move
sets from the geometric properties of the corresponding conformation-networks%
\cite{amaral,scala}. We study the conformation spaces of the homopolymers
with $m\leq 16$ on the $2D$ square lattice for different move sets. Though
the chain lengths considered in this work are relatively short, we can
construct the networks by exact enumeration. For the conformation-networks
obtained by different move sets, firstly, what are the characteristic
features of the networks? It was shown by Scala et. al.\cite{scala} that the
geometric properties of the conformation-network obtained from the mapping
of a particular conformation space are similar to those of small-world
networks. A small-world network is characterized by two properties: the
local connection is as cliquy as regular lattices, and the characteristic
path length increases logarithmically with the number of nodes\cite{ws,watts}%
. Do the conformation-networks obtained from different move sets all have
the characteristics of small-world networks? For this, we analyze the
characteristic path lengths and clustering coefficients of the networks.
Then, what are the differences in the geometric properties of the networks?
This leads us to the analyses of the distribution functions of the edge
number per node and the shortest path lengths between two nodes. Finally, we
also discuss the stability of the networks.

For the dynamical simulations of lattice polymers, the typical Monte Carlo
moves include $\left( i\right) $ end flips, $\left( ii\right) $ corner
shift, $\left( iii\right) $crankshaft move, and $\left( iv\right) $ rigid
rotation, as shown in Fig. 1. Based on these moves, we consider three
different move sets, \textbf{SA}, \textbf{SB}, and \textbf{SC}, which are
defined as the followings. The moves, $\left( i\right) $, $\left( ii\right) $%
, and $\left( iii\right) $, which change only one or two monomers in a move,
are relatively local in comparison with the move $\left( iv\right) $.
Because of the locality, these moves have been adopted very often in the
literatures\cite{sali,socci,melin}. We refer these moves as \textbf{SA}.
Note that \textbf{SA} may not satisfy the requirement of ergodicity: In the
case of $2D$ square lattice, all exactly enumerated conformations are
reachable for $m<16$, there is one unreachable for $m=16$, and the
unreachable number may become large for $m>16$\cite{chan1,chan2,hoang}. The
move $\left( iv\right) $ contains the change of more monomers, and it makes
some simple diffusive motions for groups of monomers to be possible\cite%
{skolnick1}. Since the move $\left( i\right) $ can be viewed as a
short-scale rigid rotations, we then combine the move $\left( iv\right) $
with the move $\left( i\right) $ to form the set of rotational moves, 
\textbf{SB}. Note that the move $\left( ii\right) $ or $\left( iii\right) $
from a conformation can be achieved by two or more sequential moves of 
\textbf{SB}, and all unreachable conformations for \textbf{SA} can be
obtained by the moves of \textbf{SB}. Thus, \textbf{SA} may be viewed as a
subset of \textbf{SB}. Finally, we refer the move set containing all the
moves as \textbf{SC}.

Based on the move sets \textbf{SA}, \textbf{SB}, and \textbf{SC}, we
construct the respective conformation-networks as the followings: First, we
enumerate all possible self-avoiding conformations $N\left( m\right) $ as
the nodes of the network for the chain of $m$ monomers. Note that the
degeneracy caused by the rotation and the mirror symmetry has been excluded
in $N\left( m\right) $. Then, edges exist between two nodes for which a move
of the given move set can change one to the other. The move sets, \textbf{SA}%
, \textbf{SB}, and \textbf{SC}, yield different distributions of edges among
the nodes and hence different networks. The networks can be viewed as the
folding networks in high temperature limit, all edges have the same weight.
We refer the resultant networks as \textbf{GA}, \textbf{GB}, and \textbf{GC}
for\textbf{\ }the move sets, \textbf{SA}, \textbf{SB}, and \textbf{SC},
respectively. The numbers of nodes $N\left( m\right) $ and the numbers of
edges $E\left( m\right) $ of different networks for different number of
monomers $m$ are listed in Table 1.

Firstly, we analyze the edge distributions of the networks. The number of
edges associated with a node is the number of allowed transitions from one
conformation to the others. The spread in the number of edges is
characterized by a distribution function $P\left( k\right) $ which gives the
probability for a node to have $k$ edges. Then, the average edge-number per
node is given by 
\begin{equation}
\left\langle k\right\rangle =\dsum\limits_{k}kP\left( k\right) .
\end{equation}%
The results of $\left\langle k\right\rangle $ are listed in Table 1, and
they all scale as $\left\langle k\right\rangle =a+b\log \left( N\left(
m\right) \right) $, shown in the insets of Fig. 2, with $\left( a,b\right) $
given as $\left( 3.79,0.92\right) $\ for \textbf{GA}, $\left(
3.07,2.99\right) $\ for \textbf{GB}, and $\left( 2.77,4.01\right) $\ for 
\textbf{GC}. Thus, the average edge-number per node generated by the move
set \textbf{SB }(\textbf{SC}) is about three (four) times the average number
generated by \textbf{SA}. This gives more throughway accessibility to the
native conformation and reduces the chance to be trapped in local minimum in
the folding process for the move sets \textbf{SB} and \textbf{SC}\cite%
{chan1,chan2}.

Our results of $P\left( k\right) $ versus $\Delta k=k-\left\langle
k\right\rangle $ for \textbf{GA}, \textbf{GB}, and \textbf{GC }with $m=10$, $%
12$, $14$, and $16$ are shown in Figs. 2(a), 2(b), and 2(c), respectively.
Amaral et. al. studied the subnetwork of \textbf{GA} for which, the
end-to-end distance is a parameter with a specified value for a network and
the edges between nodes are generated by the moves, corner shift and
crankshaft move\cite{amaral,scala}. Their results showed that the form of $%
P\left( k\right) $ is Gaussian. Then, we find the best fittings of the
Gaussian function,%
\begin{equation}
P\left( k\right) =\frac{1}{\sqrt{2\pi }\sigma }\exp \left[ -\frac{\left(
k-\left\langle k\right\rangle \right) ^{2}}{2\sigma ^{2}}\right] ,
\end{equation}%
for \textbf{GA}, \textbf{GB}, and \textbf{GC}, as the solid lines in Fig. 2.
Our results indicate the followings: $\left( i\right) $ For \textbf{GA}, the
distribution agrees with the Gaussian form for which, the average of the
variances of different $m$ is $\sigma _{\mathbf{GA}}=\left( 0.5748\right) 
\sqrt{N_{m}}$. $\left( ii\right) $ Comparing with the case of \textbf{GA},
there are significant deviations from the Gaussian form for the cases of 
\textbf{GB} and \textbf{GC} as shown in Figs. 2(b) and 2(c), but the
distributions are obviously not scale-free\cite{bara1,bara2,bara3}. \ \ \ \
\ \ \ 

The degree of local connections of the networks can be measured by the
clustering coefficients. We define the clustering coefficient of the node $i$
as \ 
\begin{equation}
C_{i}=\frac{2E_{k_{i}}}{k_{i}\left( k_{i}+1\right) },
\end{equation}%
where $k_{i}$ is the edge-number and $E_{k_{i}}$ is the existent edge-number
among the neighboring $k_{i}$ nodes of the node $i$. Then, the degree of
local connections of the network can be characterized by the average of the
clustering coefficients of the nodes, denoted by $C_{av}$. The values of $%
C_{av}$ for \textbf{GA}, \textbf{GB}, and \textbf{GC }with different $m$
values are listed in Table 1. The results show $%
C_{av}^{GA}>C_{av}^{GB}>C_{av}^{GC}$, and this implies that the Monte Carlo
simulation with the move set \textbf{SA} has more chances to be trapped in
some clinquy conformations comparing with those of \textbf{SB} and \textbf{SC%
}. For the network with the node-number $N$ and the average edge-number $%
\left\langle k\right\rangle $, the corresponding random network has the
average clustering coefficient $C_{av}^{ran}\approx \left\langle
k\right\rangle /N$. The results of the ratios of $C_{av}$ of \textbf{GA}, 
\textbf{GB}, and \textbf{GC} to $C_{av}^{ran}$ are shown in Fig. 3, and they
indicate that the average clustering coefficients of the
conformation-networks are much larger than that of random network.

Then, we analyze the path lengths between two nodes of the network. The
minimum number of Monte Carlo moves required for one conformation to reach
the other can be viewed as the distance between two conformations\cite%
{chan1,chan2}. Thus, the shortest path length $l$ between two nodes can be
defined as the minimum number of edges required to connect the two nodes.
Our results indicate that the distribution functions of the shortest path
lengths, $P\left( l\right) $, of different networks all agree with the
Gaussian form of Eq. (2). In Fig. 4 we plot the scaled results of the
distribution function, $\widetilde{P}\left( l\right) =\sqrt{2\pi }\sigma
P\left( l\right) $, versus $\widetilde{\Delta l}=\left( l-\left\langle
l\right\rangle \right) /\sqrt{2}\sigma $ for \textbf{GA}, \textbf{GB}, and 
\textbf{GC}, with $m=10$, $12$, $14$, and $16$. Here, we take the variances $%
\sigma $ as $\sigma _{\mathbf{GA}}=0.0489\left( m\right) ^{1.7}$, $\sigma _{%
\mathbf{GB}}=0.3057\left( m\right) ^{0.6}$, and $\sigma _{\mathbf{GC}%
}=0.5057\left( m\right) ^{0.6}$, which are determined by first finding the
least square fit to Eq. (2) to obtain $\sigma \left( m\right) $ for a given $%
m$, and then taking the average over $\sigma \left( m\right) $ of different $%
m$ to obtain $\sigma $ for a given network. The variance of $P\left(
l\right) $ for \textbf{GB} is much smaller than that for \textbf{GA}, and
this implies that the shortest distance between any two nodes does not vary
much for the networks \textbf{GB} and \textbf{GC}.

The characteristic path length of the network, $\left\langle l\right\rangle $%
, is defined as the average of the shortest path lengths for all node-pairs, 
\begin{equation}
\left\langle l\right\rangle =\dsum\limits_{l}lP\left( l\right) .
\end{equation}%
The values of $\left\langle l\right\rangle $ for \textbf{GA}, \textbf{GB},
and \textbf{GC} with different $m$ values are listed in Table 1. The results
indicate that the characteristic path length of \textbf{GB} is about half of
the length of \textbf{GA}. For the small-world networks, there exists a
cross-over size $N^{\ast }$,which is about the same order as the inverse of
the rewiring probability $p$,such that the characteristic path lengths $%
\left\langle l\right\rangle $ obey the finite-size scaling low\cite%
{barth,bar,newman}, 
\begin{equation}
\left\langle l\right\rangle =\left( N^{\ast }\right) ^{1/d}f\left( \frac{N}{%
N^{\ast }}\right) ,
\end{equation}%
where $d$ is the dimensionality of the underlie regular lattice, and $%
f\left( x\right) $ is a scaling function with the limits, $f\left( x\right)
\sim x^{1/d}$ for $x\ll 1$ and $f\left( x\right) \sim \ln x$ for $x\gg 1$.
By taking the hypothesis that the conformation network may be a small-world
network, we use the scaling form of Eq. (5) to fit the data, and the results
are shown in Fig. 5. The fittings indicate that $\left( i\right) $ the
values of $\left\langle l\right\rangle $ increase logarithmically with the
node-number $N$ for large $N$; $\left( ii\right) $ we obtain $1/d$ from the
fittings of small $N$ as $0.3427$, $0.2377$, and $0.2155$ for the networks 
\textbf{GA}, \textbf{GB}, and \textbf{GC} respectively, and then our
estimations of $d$ are $d_{GA}\sim 3$, $d_{GB}\sim 4$, and $d_{GC}\sim 4.5$;
and $\left( iii\right) $ the cross-over region $N^{\ast }\left( m\right) $
is around $m=9\sim 11$ $\left( p\sim 10^{-3}-10^{-4}\right) $ for \textbf{GA}
and $8$ $\left( p\sim 10^{-3}\right) $ for \textbf{GB} and \textbf{GC}. Note
that \ we do not take the data from $m\leq 4$ for which the node-number $N$
is less than $5$, and thence the data points available for the region of
small $N$ are few. Based on the above results, we may conclude that the
dimensionality of the conformation-space is $d\geq 3$, and the cross-over
region become narrower when the dimensionality gets larger. \ \ \ \ \ \ \ \
\ \ \ \ \ \ \ 

Finally, we analyze the ability of attack and error tolerance of the network
by studying the fragmentation caused by node-removal\cite{bara4}. The nodes
with higher degrees of connections are removed preferentially for the
analysis of attack tolerance; and the nodes are removed randomly for the
error tolerance. By removing a fraction $f$ of the nodes, we measure the
fraction of nodes contained in the largest cluster, $S$, and the average
node number, $\left\langle s\right\rangle $, contained in the fragmentary
clusters excluding the largest one. If only the removed nodes were missing
from without further breaking the largest cluster, the $S$ value decreases
from $1$ down to $0$ along the diagonal line as $f$ increases from $0$ up to 
$1$; and the $\left\langle s\right\rangle $ value remains to be one for $%
0<f\leq 1$ if the removed nodes were isolated from each other. For most
networks, we may expect that while as the $S$ values start to decrease more
rapidly than the diagonal line at some fraction $f_{m}$, and drop to zero at
the critical fraction $f_{c}$; the $\left\langle s\right\rangle $ value
start to increase more rapidly from $\left\langle s\right\rangle =1$ at $%
f_{m}$, and reach the maximum at $f_{c}$. The results of $S$ and $%
\left\langle s\right\rangle $ as function of $f$ are shown in Fig. 6 for the
networks \textbf{GA}, \textbf{GB}, and \textbf{GC} with $m=16$. Our results
show that the $f_{c}$ value\ is very closed to $1$, and the stability of the
networks is very analogous to random networks. \ \ \ \ 

In summary, we divide the frequently used Monte Carlo moves into three
different move sets, and construct the corresponding conformation-networks.
The networks all have the characteristics of small- world: $\left( i\right) $
the local neighborhood is more cliquy than that of random networks, and $%
\left( ii\right) $ the characteristic path length increases logarithmically
with the number of nodes. The dimensionalities of the conformation-spaces
are $d\geq 3$. Our analyses also indicate that the networks are as robust as
random graphs. Among different Monte Carlo moves, the rigid rotation has
drastic effect on the geometric properties of the network: $\left( i\right) $
it renders the connection distribution to be non-Gaussian, and $\left(
ii\right) $ it reduces greatly the characteristic path length. Thus, the
Monte Carlo move, rigid rotation, may change the folding kinetics
significantly from that of the local moves, corner shift and crankshaft
move. \ \ \ \ \ \ 

\bigskip \textbf{Acknowledgments}

This work was partially supported by the National Science Council of
Republic of China (Taiwan) under the Grant No. NSC93-2212-M033-005. We thank
National Center for High-performance Computing and the Computing Centre of
Academia Sinica for providing \ the computation facilities.

\newpage

\FRAME{ftbpFU}{4.5359in}{2.5227in}{0pt}{\Qcb{Examples of typical Monte Carlo
moves: $(a)$ end flips, $\left( b\right) $ corner shift, $\left( c\right) $%
crankshaft move, and $\left( d\right) $ rigid rotation. The current
conformation is shown in thick lines, and possible new conformations are
shown in broken lines.}}{}{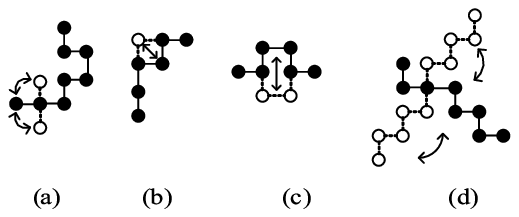}{\special{language "Scientific
Word";type "GRAPHIC";display "USEDEF";valid_file "F";width 4.5359in;height
2.5227in;depth 0pt;original-width 2.4837in;original-height 1.228in;cropleft
"0";croptop "1";cropright "1";cropbottom "0";filename
'fig1.eps';file-properties "XNPEU";}}

\FRAME{ftbpFU}{4.2229in}{3.3849in}{0pt}{\Qcb{The probability distribution of
edges, $P(k)$, versus $\Delta k=k-\left\langle k\right\rangle $ for the
networks $\left( a\right) $\textbf{GA}, $\left( b\right) $\textbf{GB}, and $%
\left( c\right) $\textbf{GC}. Here, $\left\langle k\right\rangle $ is the
average edge number per node, and the solid lines are the best fittings of
the Gaussian function given in the text. For each network, $\left\langle
k\right\rangle $ versus $\log \left( N\left( m\right) \right) $ for the node
number $N\left( m\right) $ with the monomer number $m$ ranged from $8$ to $%
16 $ is shown in the inset, and the straight solid line corresponds to the
relation $\left\langle k\right\rangle =a+b\log \left( N\left( m\right)
\right) $ with the values of $a$ and $b$ given in the text. \ }}{}{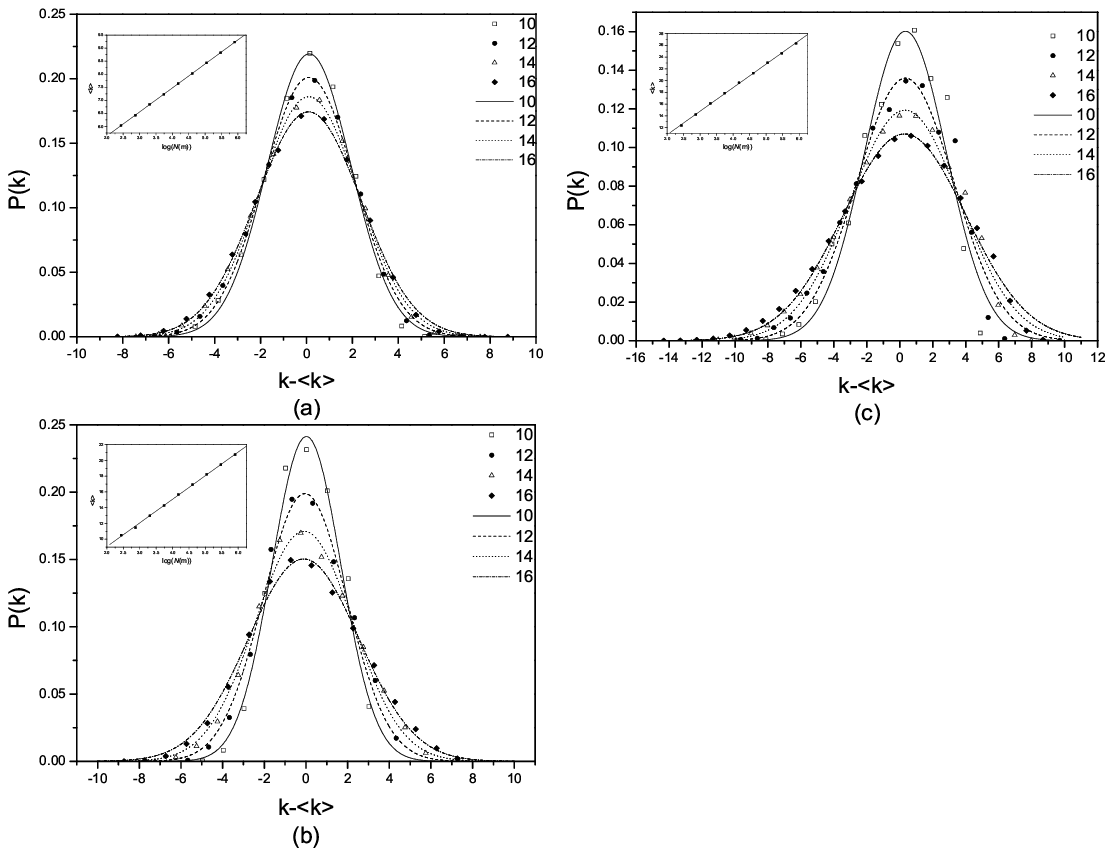}{%
\special{language "Scientific Word";type "GRAPHIC";maintain-aspect-ratio
TRUE;display "USEDEF";valid_file "F";width 4.2229in;height 3.3849in;depth
0pt;original-width 5.2511in;original-height 4.203in;cropleft "0";croptop
"1";cropright "1";cropbottom "0";filename 'fig2.eps';file-properties
"XNPEU";}}

\FRAME{ftbpFU}{4.158in}{3.346in}{0pt}{\Qcb{The ratios of the average
clustering coefficients, $C_{av}$, of the networks \textbf{GA}, \textbf{GB},
and \textbf{GC} to the average clustering coefficients of random lattices $%
C_{av}^{ran}$ versus $\log \left( N\right) $ with the node number $N\left(
m\right) $ and the monomer number $m$ ranged from $8$ to $16$.}}{}{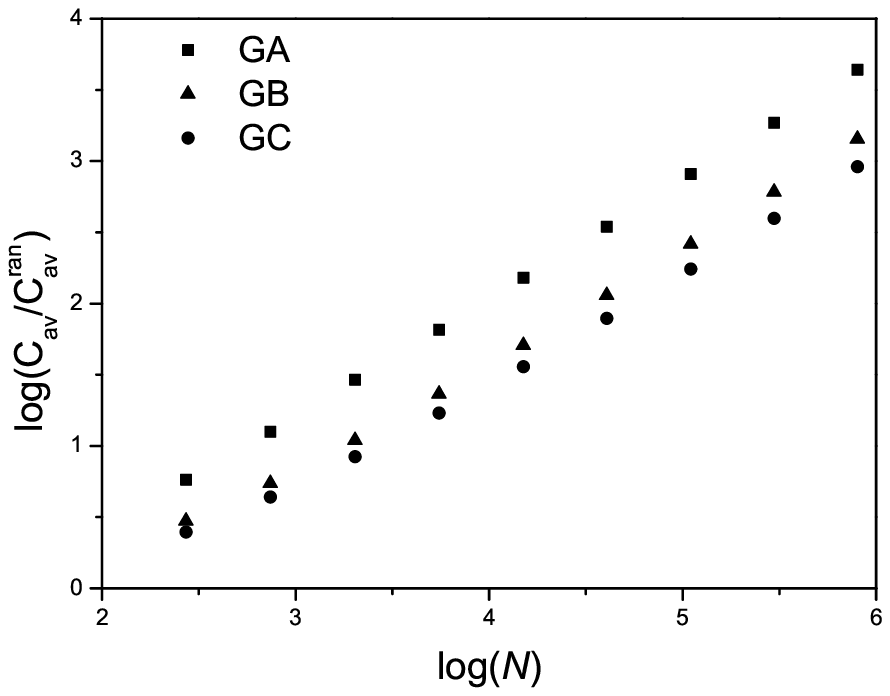}{%
\special{language "Scientific Word";type "GRAPHIC";maintain-aspect-ratio
TRUE;display "USEDEF";valid_file "F";width 4.158in;height 3.346in;depth
0pt;original-width 4.19in;original-height 3.3684in;cropleft "0";croptop
"1";cropright "1";cropbottom "0";filename 'fig3.eps';file-properties
"XNPEU";}}

\FRAME{ftbpFU}{4.3102in}{3.3762in}{0pt}{\Qcb{ The scaled result of the
distribution function of the shortest path lengths $P\left( l\right) $, $%
\widetilde{P}\left( l\right) =\protect\sqrt{2\protect\pi }\protect\sigma %
P\left( l\right) $, versus $\widetilde{\Delta l}=\left( l-\left\langle
l\right\rangle \right) /\protect\sqrt{2}\protect\sigma $ for $\left(
a\right) $ \textbf{GA}, $\left( b\right) $ \textbf{GB}, and $\left( c\right) 
$\textbf{GC}, with $m=10$, $12$, $14$, and $16$. The averages of the
shortest path lengths for all node-pairs $\left\langle l\right\rangle $ are
given in Table 1, and the variances $\protect\sigma $ are $\protect\sigma _{%
\mathbf{GA}}=0.0489\left( m\right) ^{1.7}$, $\protect\sigma _{\mathbf{GB}%
}=0.3057\left( m\right) ^{0.6}$, and $\protect\sigma _{\mathbf{GC}%
}=0.5057\left( m\right) ^{0.6}$. The solid lines are the results of the
Gaussian function given in the text.}}{}{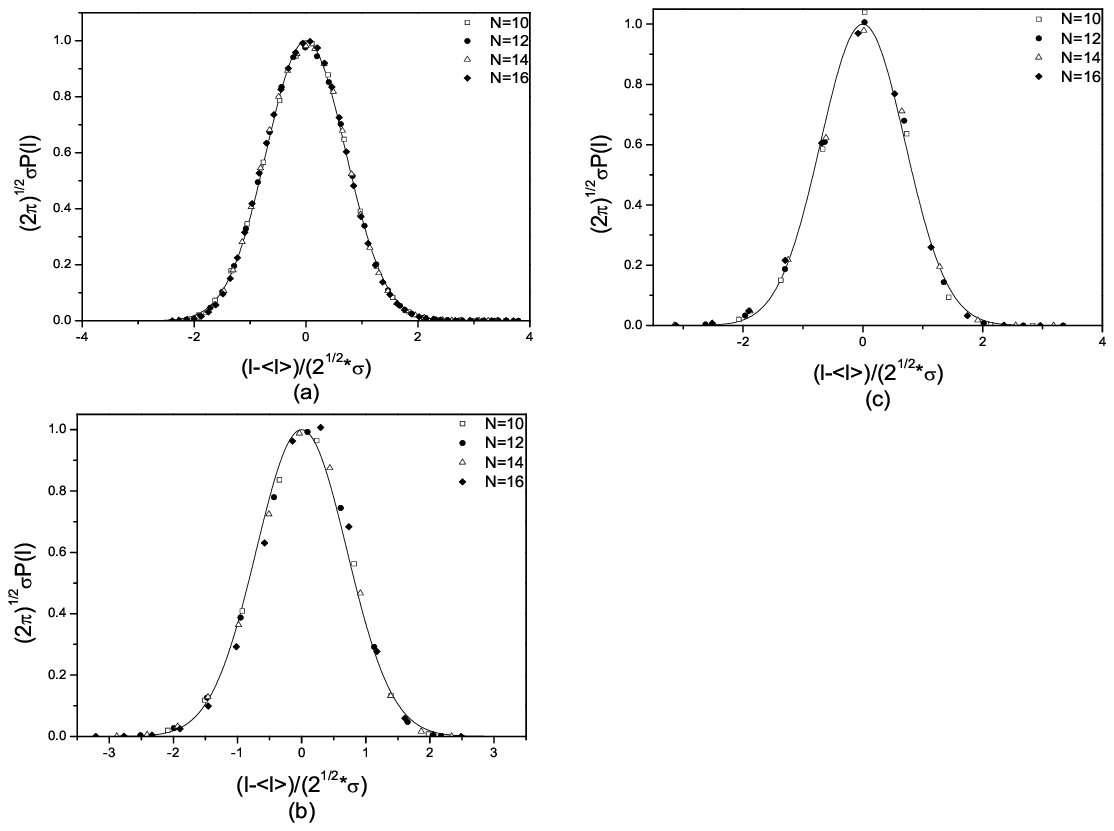}{\special{language
"Scientific Word";type "GRAPHIC";maintain-aspect-ratio TRUE;display
"USEDEF";valid_file "F";width 4.3102in;height 3.3762in;depth
0pt;original-width 5.2511in;original-height 4.1053in;cropleft "0";croptop
"1";cropright "1";cropbottom "0";filename 'fig4.eps';file-properties
"XNPEU";}}

\FRAME{ftbpFU}{4.35in}{3.4368in}{0pt}{\Qcb{The characteristic path length $%
\left\langle l\right\rangle $ versus the logarithm of the node-number $N$, $%
\log \left( N\right) $, for the networks $\left( a\right) $ \textbf{GA}, $%
\left( b\right) $ \textbf{GB}, and $\left( c\right) $ \textbf{GC }with the
monomer number $m$ ranged from $5$ to $16$. The insets are the plots of log$%
\left( \left\langle l\right\rangle \right) $ versus $\log \left( N\right) $
for the same data. The solid lines are the results of the limiting scaling
forms given in the text.}}{}{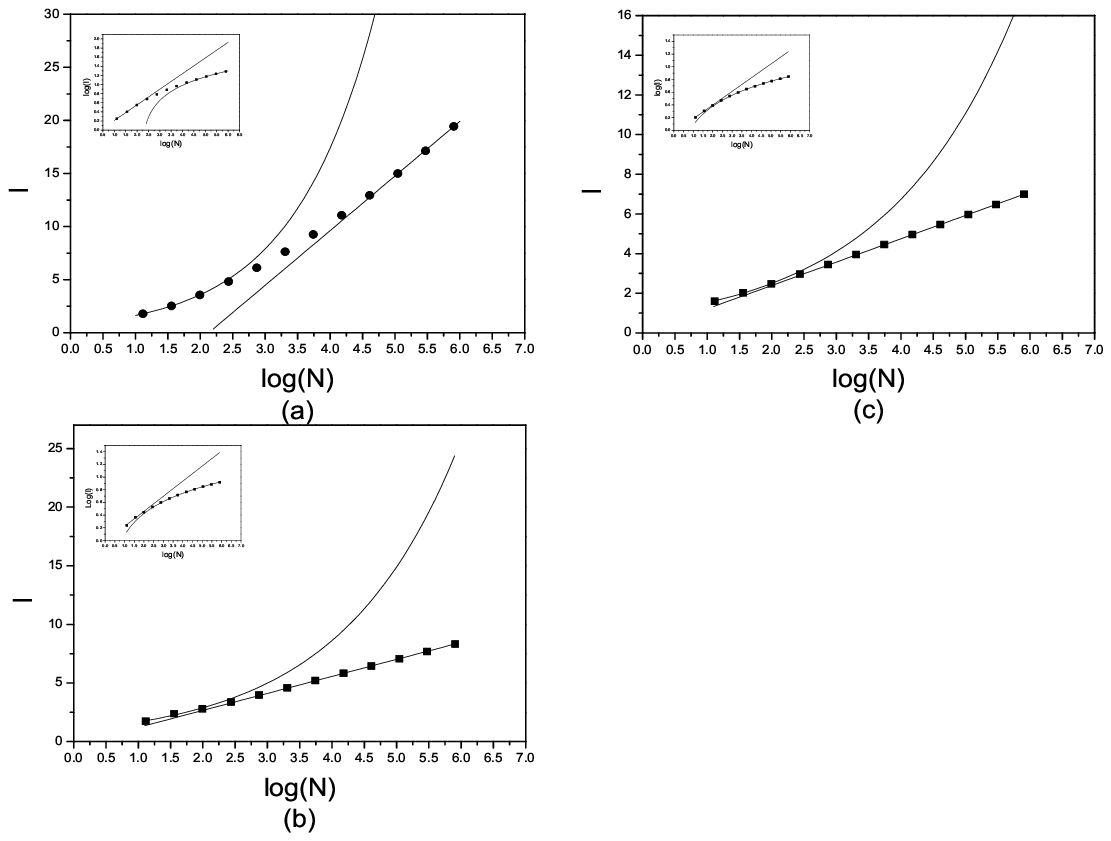}{\special{language "Scientific
Word";type "GRAPHIC";maintain-aspect-ratio TRUE;display "USEDEF";valid_file
"F";width 4.35in;height 3.4368in;depth 0pt;original-width
5.2209in;original-height 4.1208in;cropleft "0";croptop "1";cropright
"1";cropbottom "0";filename 'fig5.eps';file-properties "XNPEU";}}

\FRAME{ftbpFU}{4.5031in}{2.4215in}{0pt}{\Qcb{The fraction of nodes contained
in the largest cluster, $S$, and the average node number, $\left\langle
s\right\rangle $, contained in the fragmentary clusters excluding the
largest one versus the fraction $f$ of the nodes removed for $\left(
a\right) $ attack and $\left( b\right) $ error tolerance of the networks 
\textbf{GA}, \textbf{GB}, and \textbf{GC} with $m=16$. \ \ \ \ }}{}{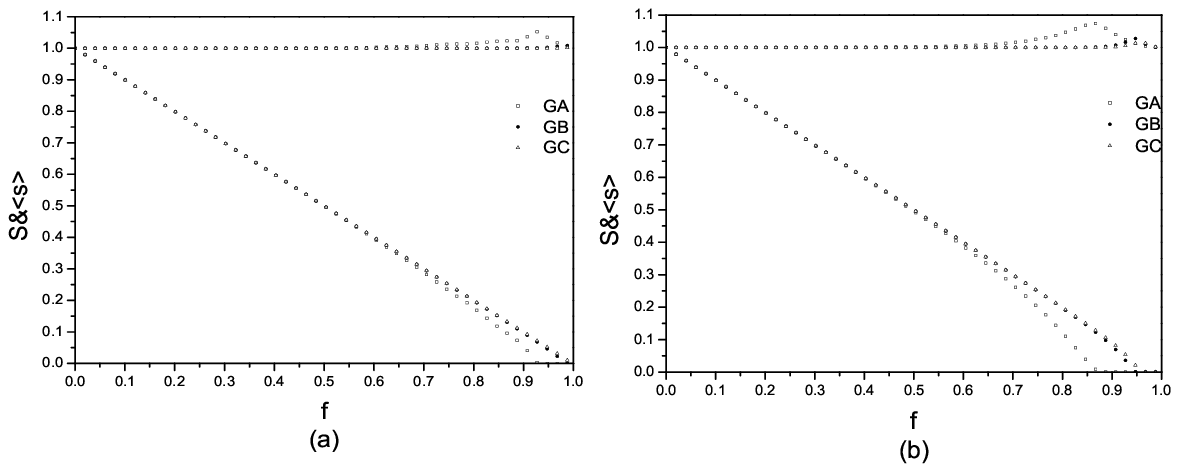%
}{\special{language "Scientific Word";type "GRAPHIC";display
"USEDEF";valid_file "F";width 4.5031in;height 2.4215in;depth
0pt;original-width 5.5287in;original-height 2.6844in;cropleft "0";croptop
"1";cropright "1";cropbottom "0";filename 'fig6.eps';file-properties
"XNPEU";}}

\bigskip

\begin{table}[tbh]
\caption{ Various geometric quantities of the conformation-networks \textbf{%
GA}, \textbf{GB}, and \textbf{GC} with different number of monomers $m$: the
numbers of nodes $N$, the numbers of edges $E$, the average edge number per
node $\left\langle k\right\rangle $, and the characteristic path length $%
\left\langle l\right\rangle $.}%
\begin{tabular}{ccccc}
\hline
$m$ & 10 & 12 & 14 & 16 \\ \hline
$N$ & 2034 & 15037 & 110188 & 802075 \\ 
$E^{GA}$ & 6966 & 57451 & 464687 & 3702485 \\ 
$E^{GB}$ & 13194 & 117839 & 1005304 & 8314161 \\ 
$E^{GC}$ & 16397 & 147673 & 1268544 & 10554679 \\ 
$\left\langle k\right\rangle ^{GA}$ & 6.8496 & 7.6413 & 8.4344 & 9.2323 \\ 
$\left\langle k\right\rangle ^{GB}$ & 12.9735 & 15.6732 & 18.2471 & 20.7316
\\ 
$\left\langle k\right\rangle ^{GC}$ & 16.1229 & 19.6413 & 23.0251 & 26.3184
\\ 
$\left\langle l\right\rangle ^{GA}$ & 7.6369 & 11.0731 & 15.0046 & 19.4403
\\ 
$\left\langle l\right\rangle ^{GB}$ & 4.5953 & 5.8286 & 7.0726 & 8.3236 \\ 
$\left\langle l\right\rangle ^{GC}$ & 3.9555 & 4.9611 & 5.9723 & 6.9869 \\ 
\hline
\end{tabular}%
\newline
\end{table}

\end{document}